\documentclass{cta-author}

{}
{}
{}

\usepackage{breqn}
\usepackage{adjustbox}
\DeclareUnicodeCharacter{FF0C}{\textvisiblespace}

\usepackage[english]{babel}
\usepackage[utf8]{inputenc}
\usepackage{algorithm}
\usepackage[noend]{algpseudocode}
\usepackage{amsfonts}
\usepackage{url}
\usepackage{listings}
\usepackage{fixltx2e}
\usepackage{amsmath}
\usepackage[
	n,
	operators,
	advantage,
	sets,
	adversary,
	landau,
	probability,
	notions,	
	logic,
	ff,
	mm,
	primitives,
	events,
	complexity,
	asymptotics,
	keys]{cryptocode}
\usepackage{tabto}
\usepackage{array}

\begin{document}

\title{ 
Enhanced Cyber-Physical Security Using Attack-resistant Cyber Nodes and Event-triggered Moving Target Defence}

\author{\au{Martin Higgins\textsuperscript{1}, }
\au{Keith Mayes\textsuperscript{2}, }
and 
\au{Fei Teng\textsuperscript{1*}}
}

\address{\add{1}{Imperial College London, Department of Electrical and Electronic Engineering, South Kensington, London, UK}
\add{2}{Royal Holloway, University of London, Information Security Group, Egham, London, UK}
\email{f.teng@imperial.ac.uk}}

\begin{abstract}
This paper outlines a cyber-physical authentication strategy to protect power system infrastructure against false data injection (FDI) attacks. We demonstrate that it is feasible to use small, low-cost, yet highly attack-resistant security chips as measurement nodes, enhanced with an event-triggered moving target defence (MTD), to offer effective cyber-physical security. At the cyber layer, the proposed solution is based on the MULTOS Trust-Anchor chip, using an authenticated encryption protocol, offering cryptographically protected and chained reports at up to 12/s. The availability of the trust-anchors, allows the grid controller to delegate aspects of passive anomaly detection, supporting local as well as central alarms. In this context, a distributed event-triggered MTD protocol is implemented at the physical layer to complement cyber side enhancement. This protocol applies a distributed anomaly detection scheme based on Holt-Winters seasonal forecasting in combination with MTD implemented via inductance perturbation. The scheme is shown to be effective at preventing or detecting a wide range of attacks against power system measurement system.
\end{abstract}

\maketitle




%

\section{Introduction}
\label{sec:intro}
The power system is arguably the most critical of all modern networked infrastructures. To some extent, water, communications, sanitation and defence systems are dependent on a stable electricity supply and it is no wonder that the power system has become a key target for hackers seeking to disrupt or destroy. In 2015, Ukraine experienced one of the first successful (and well documented) cyber attack against a distribution power grid with the consequence of around 300,000 people disconnected from their power supply \cite{Liang2017TheAttacks}. This attack brought significant focus to the area of cyber-attacks against power systems. However, to some extent, the impact of the Ukraine attacks was limited as the attackers chose a non-stealthy method of implementing the attack. If the attackers  had  opted  for  a stealthy  attack type, such as  False Data Injection (FDI) attacks, the attackers may have been able to continue attacking and achieve much greater damages. 

In the past, the majority of works focused on either physical or cyber security enhancements in isolation however, more recently, works have been published which demonstrate the importance of considering overlapping cyber and physical solutions \cite{He2016Cyber-physicalSurvey}. In this paper we propose and evaluate a practical and affordable system design that protects against attacks against both the physical and cyber layers and allows for distributed protection.

The cyber domain of power system is susceptible to various attacks. National guidelines suggest the use of cryptographic tools  for authentication and encryption \cite{Pillitteri2014GuidelinesSecurity}. However, as outlined in \cite{Ghosh2019AChallenges}, supervisory control and data acquisition (SCADA) networks (which are used for monitoring and controlling power systems), often lack secure modern encryption. This can be due to the age of assets (encryption and security of networks being a more recent concern) or requirements for high speed in data processing with encryption adding too much time delay in measurement transmission \cite{Tan2017SurveyApproach}. The size and scale of power networks can also makes the cost of refitting prohibitive with increased focus put on 'sweating' assets rather than incorporating new infrastructure. As a result, while new cyber layer protection solution is urgently required, it must be low cost (to ensure practical take up), secure by the Common Criteria security framework \cite{2017CommonModel} and shown not to impede data transmission. 

Furthermore, the links between cyber and physical systems are also vulnerable to the attacks. In the  case of the power measurement system, a simple alternative to attacking the cyber-secured measurement node, is to modify the analogue (or physical) measurement source, which can not be protected by cyber domain solutions. The attackers may conduct FDI attacks by manipulating analogue measurements before they reach cyber-secured point. 

One attack type that uses this form of measurement changing is the false data injection (FDI) attack. FDI attacks use system information to remain hidden while attacking systems measurement to misguide the WLS-based state estimation \cite{Monticelli1999StateApproach}. Although extensive research have been conducted to detect FDI attack, the vast majority of papers operate on the assumption of residual testing based on centralised state estimation (CSE) \cite{Deng2017FalseSurvey}. The consequences of a successful FDI attack can be severe. For example, in \cite{Tan2018Cyber-attackModel} it is shown that FDI attacks can be used to promote load shedding via the system operator or overload certain lines and in \cite{Chung2018LocalGrid} it was shown how FDI attacks can be used to mask transmission outages. A few papers recently started to tackle FDI attacks from a distributed perspective. In \cite{Yang2016OnEstimation},  \cite{Gabbar2017OnGrids} \& \cite{Karimipour2017RobustCyber-Attack} FDI attacks against distributed Kalman style or extended Kalman style filtering are explored to show how dynamic style estimation (at the individual state or measurement) can still have potential susceptibilities to attack. We consider that to be secure at the physical layer the system must be protected against these types of attacks. Topology-driven moving target defence (MTD) has been shown previously to be effective against FDI attack. It is possible to implement MTD in the physical power system either through both transmission switching \cite{Wang2015EffectsNetworks} or more commonly by admittance perturbation via distributed flexible AC transmissions (D-FACTs) devices \cite{Morrow2012TopologyInjection} \cite{Liu2017ReactanceEstimation} \cite{Li2019OnDevices} \cite{Tian2019EnhancedGrids}. Very few papers have explored transmission switching as a legitimate method of applying of MTD in the power system. This is because, while in theory it could be effective at driving increased residuals in the case of an attacker unaware of MTD as we showed in \cite{Higgins2020StealthySystems}, the impacts of such an approach on power system control and cost to operation would be undesirable. Also, from the attackers perspective identifying these changes and countering are much easier in the case of transmission switching. By comparison, D-FACTS based applications are much more common and cause less overall impact to power system stability and have lower operational costs. Indeed, while it has been shown that MTD can be effective, limited work has been put into how it should be implemented in power systems. Questions around the frequency and when MTD should be applied remain. Continuous application of MTD will result in significantly higher costs \cite{Lakshminarayana2018Cost-BenefitGrids} and time-based application i.e. once per hour, is possible but will risk attacks slipping between applications. In this paper, we argue that MTD should only be applied if there is a creditable suspicion that the system is under attack. In \cite{Tamba2019AnControl} an event driven MTD is proposed for control systems and we adopt a similar idea for power system application.

\begin{figure*}[t]
\centering
\includegraphics[height=10cm]{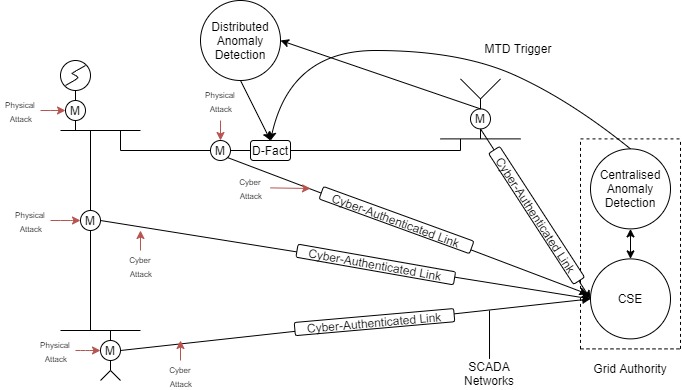} 
\caption{Outline of proposed cyber-physical defence model featuring the MTD trigger mechanism, metering and distributed anomaly detection.}
\label{fig:physicaldefence}
\end{figure*}

\subsection{Contributions}
This paper proposes and evaluates a practical and affordable system design that protects against both physical and cyber layer attacks and allows for distributed protection. We demonstrate how a cyber-physical authentication process can provide a foundation for system security at every level of measurement and transmission. This process is illustrated in Fig. \ref{fig:physicaldefence}.
\subsubsection{Cyber Layer}

In the cyber layer, we propose creating trustworthy measurement nodes by using a small pay load, encrypted protocol to secure the measurement reporting system while maintaining the required frequency of reporting. Each node is based on a highly tamper-resistant security chip that performs measurement capture, processing and reporting. The chip acts as a trustworthy platform, supporting the secure communication protocol with the grid authority (GA). The combination of the chip and protocol, offers effective protection against implementation attacks, and communications attacks, such as man-in-the-middle/FDI. The secured platform also allows confident delegation of some limited processing to the nodes, such as the proposed anomaly check and MTD triggering. The MULTOS Trust-Anchor for secure measurement transmission and MTD delegation has not been applied before. This maybe because MULTOS platforms have been traditionally in the form of secured smart card and passport chips, which do not have general external I/O. Although Authenticated Encryption has been standardised for some time; it has only relatively recently been studied on such security chips. The design secures the logical and implementation security of the system, but importantly, it provides a framework for delegated/distributed implementation of trusted functions and self-defensive mechanisms. We have also shown that the secure nodes can sample, process and locally react, much faster than would be feasible for a centralised system, offering the potential for new algorithms and research.

\subsubsection{Physical Layer}

In the physical layer, we propose a protection protocol based on event-triggered MTD. The protocol consists of an initial anomaly detection followed by MTD and traditional CSE. The anomaly detection uses Holt-Winters seasonal forecasting distributed to the individual measurement. The seasonality captures the intra-day demand differences to minimise the overall window for attack. If the anomaly detection is triggered, MTD is then implemented via inductance perturbation of D-FACTS devices. This changes the system model and drive detection of the FDI attack via increased residual errors. The event trigger comes prior to the application of MTD and is based on distributed measurements to target and minimise the overall use of MTD. This is important as the FDI attacks are stealthy from a centralised residual detection perspective, but their impacts can still be seen in the distributed changes to power flows. One of the key motivation drivers for this research is that MTD can be expensive to implement given the sub-optimal use of assets from an optimal power flow perspective as shown in \cite{Lakshminarayana2020Cost-BenefitGrids}. Very few works have addressed the importance of distributed measurement with respect to FDI attacks and none that we have seen combine a distributed and centralised approach as we have done here. 

This paper proceeds as follows: firstly Section \ref{sec:background} provides background to the proposed approaches. This includes outlining the state estimation environment for the power system. Section \ref{sec:solution} proposes and describes a secure solution based on an Authenticated Encryption (AE) protocol and the use of distributed Security Chips (SC). A novel event-triggered MTD is proposed in \ref{sec:event} and the Holt-Winters seasonal forecasting technique as triggering strategy outlined. Experiments, performance results and the effectiveness of security measures are discussed in Section \ref{sec:exp}, with conclusions and suggestions for future work in Section \ref{sec:con}.

\section{Background}
\label{sec:background}
\subsection{Cyber Vulnerability of SCADA System in Power System}

\label{sec:secreq}
An attacker has several options when seeking to undermine the measurement and reporting system. The node itself can be attacked to either ensure that the node does not perform as intended or to reveal a credential (such as a cryptographic key) or proprietary functionality. Also possible, is that the measurement source (e.g. value to be sampled) could be modified so that incorrect values are captured, but then securely delivered to the GA (upstream physical attack). However, the node's remote communications link presents the most attractive target, and attackers may seek to remove, modify or replay valid transmissions, or generate fake transmissions that might be accepted by the node, and inject consequently false data into the state-estimator downstream of the physical system. 
\subsubsection{Node Implementation Attacks}
Critical infrastructure may be attacked in a variety of ways. Adhering to information security best practice for the functional design of a measurement node, (e.g., for algorithm, protocols, processes and keys) is not sufficient, as attacks may target the implementation. 

Logical security attacks, target weak design typically in the algorithms, keys and poor software within either the node platform, the measurement application, or the loading and configuration functionality. Physical attacks, generally require considerable expertise, equipment and time. For example, an attempt may involve decapsulating a chip, hardware reverse engineering, probing buses and memories and modifying tracks. Fault attacks are less intrusive, they disrupt normal operation of the chip; but without damage. For example, faults can be generated by, voltage glitches or radiation pulses. Under fault condition, an unprotected chip may reveal sensitive information, for example RSA keys \cite{Boneh1997OnComputations}. Side-channel leakage, is the leakage of sensitive information via an unintended channel. This can be key or data-dependent timing, variations in power consumption or electromagnetic emissions. Simple analysis techniques  (see \cite{Kocher1996TimingSystems} \cite{Kocher1999DifferentialAnalysis}) are powerful against naive implementations. In the case of a power measurement system, a simple alternative to attacking the measurement node, is to modify the analogue measurement source (physical attack). A similar effect is achieved by manipulating any part of the path that digitises and communicates the source measurement value, up until the point when it is within the node.

All of the these attacks can be successful against implementations that do not have specialist security protection. However, there is much industry expertise for the protection of secured micro-controller chips, which we exploit in our proposal, as discussed in Section \ref{sec:solution}.

 \subsubsection{Adversary Threat Model}
We make the assumption of a well resourced attacker, similar to those seen in \cite{Liang2017TheAttacks} capable of launching attacks in both the cyber and physical realms. The attacker can attack either remotely via SCADA networks or directly at the physical sensor locations themselves.  We also assume that the attacker has the capability to structure an FDI attack via the changing of meter measurements so that it stealthy. Consistent with this is the assumption that the attacker has knowledge of the original system topology of the power system, but is not aware of MTD configurations. We now proceed to outline the respective cyber and physical attack types in this section.

 \subsubsection{Node Communication Attacks}
To undermine communications between two legitimate parties, an attacker may attempt to:

\begin{enumerate}
  \item Passively eavesdrop
  \item Send a new message assuming a legitimate identity
  \item Delay, replay, or re-order legitimate messages
  \item Modify a legitimate message
  \item Block some legitimate messages
  \item Denial of Service (DoS) transmissions
\end{enumerate}

The goal of eavesdropping is to capture unprotected data or information that can reveal credentials, such as keys or counters, which aid attacks against the protection. For grid measurement transmission, we are primarily concerned with the integrity of measurements, rather than confidentiality, however encrypting messages also protects associated control data and provides an extra barrier to attackers.
Depending on the communications channel, an attacker might simply be able to source a fake message, or may have to be positioned as a "man-in-the-middle". In either case, the attack will be defeated by mutual authentication of communicating parties, as the attacker will not have the cryptographic credentials to authenticate, or to fake, or modify a valid message.
Most communications systems experience some variation in transmission delay, but, excessive delay (or replay) can be detected by timestamps. Message re-ordering can also be detected by chaining encrypted transmission sequences. If in normal operation, a system occasionally loses messages, then an attacker could block some without being detected, however limited data loss could be overcome by redundant transmissions.
DoS would prevent legitimate messages getting through, however it is a condition, detectable by the legitimate communicating parties; which should be able to take some alarm and/or remedial action.

\subsection{False Data Injection Attacks on State Estimation}

In terms of the physical attack target, we consider a power grid state estimation system subject to FDI style attacks \cite{Liu2011FalseGrids}. FDI attacks focus on misleading system operators by introducing false information within the distributed grid system measurements, to damage the physical grid, or its operation, or cause financial consequences. From the attackers perspective, a major advantage of the FDI attack is that it may be performed stealthily. Meaning that if the false data is structured to seem plausible it will appear like a system under normal conditions to the operator. While the majority of early works focused on a linear approximation of the FDI attack, it was shown in  \cite{Hug2012VulnerabilityCyber-attacks} that an AC attack model was indeed possible provided the attacker had knowledge of the system topology. A comprehensive review of FDI attacks can be found in \cite{Deng2017FalseSurvey}. 
We outline here the basis for CSE  and FDI attacks. Under CSE, we consider a set of $n$ state variables $\textbf{x} \in \mathbb{R}^{n\times1}$ estimated by analysing a set of $m$ meter measurements $\textbf{z} \in \mathbb{R}^{m\times1}$ with a corresponding error vector $\textbf{e} \in \mathbb{R}^{m\times1}$ . The non-linear vector function $\textbf{h}(\textbf{.})$ relating meter measurements $\textbf{z}$ to states $\textbf{h}(\textbf{x}) = (h_1(\textbf{x}),h_2(\textbf{x}),...,h_m(\textbf{x}))^T$ is shown by 

\begin{equation}
    \textbf{z} = \textbf{h(x)} + \textbf{e}.
    \label{generalized state equation}
\end{equation}

With active and reactive power flow measurements under the non-linear expression defined by

\begin{equation}
    P_{ij} = V_{i}^2 g_{ij} - V_{i}V_{j}g_{ij}\cos{\Delta \theta_{ij}}-V_{i}V_{j}b_{ij}\sin{\Delta \theta_{ij}},
   \label{Pac1}
\end{equation}

\begin{dmath}
    Q_{ij} = -V_{i}^2(b_{ij}+b_{ij}^{sh}) + V_{i}V_{j}g_{ij}\cos{\Delta \theta_{ij}}-V_{i}V_{j}b_{ij}\sin{\Delta \theta_{ij}}.
    \label{Qac}
\end{dmath}

While the linear model is no doubt easier, it contains a number of assumptions around voltage angle and other areas \cite{Wood2014PowerControl}. Under certain circumstances, a linear model can be used for operation due to it's close approximation. These assumptions are:

\begin{enumerate}
    \item Line resistances are considered negligible compared to line reactance i.e. $R_L<<X_L$
    \item The voltage profile is flat and voltage amplitude equivalent across all nodes
    \item The voltage angle differences are small which results in a linearisation of the sine/cosine elements in power flow equations  
\end{enumerate}

Under these assumptions $V_i \approx V_j$, the small angle differences result in $\sin{\Delta\theta_{ij} \approx \theta_{ij}}$ and $g_{ji}$ is much smaller than $b_{ji}$, leading to the linear approximation for power flow

\begin{equation}
    P_{ij} = -b_{ij}\theta_{ij}.
    \label{Pdc}
\end{equation}

Which gives the linear approximation as

\begin{equation}
    \textbf{z} = \textbf{Hx} + \textbf{e}.
\end{equation}

The state estimation problem is to find the best fit estimate of $\hat{\textbf{x}}$ corresponding to the measured power flow values of $\textbf{z}$. Under the most widely used estimation approach, the state variables are determined by minimization of a WLS optimization problem as 

\begin{equation} \label{chisquaredfull}
 {\mathrm{min}}\,J(\textbf{x}) =  (\textbf{z}-\textbf{h(x)})^T\textbf{W}(\textbf{z}-\textbf{h(x)}).
\end{equation}
\textbf{W} is a diagonal $m \times m$ matrix consisting of the measurement weights. Under the AC model, the minimisation will be done using the partial derivative with respect to the states and iterative processes.

This is done using iterative processes usually the Newton-Raphson \cite{Monticelli1999StateApproach} utilising the Jacobian  $\textbf{J}$ of partial derivatives

$$
\textbf{J}=
\begin{vmatrix}
\frac{\delta h_1}{\delta x_1}&...&\frac{\delta h_1}{\delta x_n}\\
...&...&...\\
\frac{\delta h_1}{\delta x_m}&...&\frac{\delta h_m}{\delta x_n}\\
\end{vmatrix}.
$$

The current approach in power systems operation for bad data detection (BDD) is to use the 2-norm of the measurement residual with a detection threshold $\eta$ \cite{Monticelli1999StateApproach}.

\begin{equation} \label{residual}
    r =  ||\textbf{z} -\textbf{h}(\hat{\textbf{x}})||_2.
\end{equation}

If $r_t>\eta$, BDD alarms will trigger and the system operator will discard the result. However, recent papers have shown these methods can be attacked by structuring the attack vector in terms of the system topology

\begin{equation} \label{residual}
  \textbf{z}_a = \textbf{h}(\textbf{x}+\textbf{c})
\end{equation}
 where $c$ is a desired attack bias of length $n$ applied to the state angles or voltage measurements. 

\begin{equation} \label{residual}
    r =  ||\textbf{z}_a -\textbf{h}(\hat{\textbf{x}+\textbf{c}})||_2.
\end{equation}

As discussed, the $\textbf{z}_a$ vector is designed considering the state set $\textbf{x}$ and bias $\textbf{c}$ whilst also considering the transformation function $\textbf{h}()$. This ensures that the residual can be designed such that it will fall within acceptable limits and ensure stealthiness.

\section{MULTOS Trust-Anchor }
\label{sec:solution}
Our proposed solution requires trustworthy measurement nodes that are \textit{strongly} attack resistant. Fortunately, there is a well-established, process of assessment to determine this, known as Common Criteria (CC) \cite{2017CommonModel}. The CC evaluation is most commonly applied to security chips such as used in bank cards, passports or Hardware Security Modules (HSM). High levels of evaluation, confirm strong resistance to all known attacks, including, logical, physical, fault and side-channel. The resistance is based on specialist hardware, supported by software defensive measures.
A CC evaluated SC should not be vulnerable to logic flow attacks, being compliant with security best-practices for design, and loading. For defending against physical attack, the CC chips will have numerous effective defences including: passive/active shields (to impede probing), hardware encrypted buses and memories (to prevent data discovery), light and anomaly sensors (to detect decapsulation and faults) and scrambled circuitry (to impede reverse engineering). For fault attack resistance, the hardware sensors, detect fault insertion and prevent a response useful to the attacker. Software countermeasures provide secondary defences; e.g., verifying an algorithm result before providing an output, and redundantly testing flags and loop counts. Side-channel leakage, is well defended against, with countermeasures that impede statistical averaging of signals, or reduce the source generation of the leakage. Attack countermeasures include, power smoothing, noise insertion, randomisation of execution, timing equalisation and dual-rail logic (or dual CPUs).

MULTOS \cite{MultosInc2020MultosWebsiteb} chip platforms are quite common in CC evaluations, and the MULTOS \cite{MultosInc2020MULTOSWebsite} Trust-Anchor was selected for our proposal. MULTOS is a high security multi-application smart card Operating System (OS) that is managed by the MULTOS Consortium. MULTOS chips are found in  range of products including payment smart cards and electronic passports, achieving high-levels of CC certification. The initialisation, and personalisation of these products has strict security controls, meaning that only reviewed code and data from certified developers can be loaded. An introduction to MULTOS can be found in \cite{Mayes2017Smart17}. The Trust-Anchor is intended as a deployed hardware security model (HSM) for the Internet of Things (IoT) and differs from a smart card chip, in that it has added I/O capability and free-runs. Despite strong defensive capabilities, the chips lag behind the state-of-the-art in CPU performance and memory sizes, although the chips excel in cryptographic operations as they incorporate relatively high-speed Crypto Co-processor (CCoP) hardware; related performance evaluations can be found in \cite{MayesPerformanceAlgorithm} \cite{Mayes2017PerformanceCo-processors} \cite{Mayes2020PerformanceTrust-Anchor}. The choice of SC, means we can reasonably assume that physical, fault, side-channel and malware attacks are \textit{practically infeasible}, and so our focus turns towards a protocol for securing the interaction between the SC and GA. A previous study for EMVCo \cite{EMVcoLLC1999EMVcoWebsite} measured the performance of Authenticated Encryption (AE) \cite{IEC200919772Encryption} for future payment card processing; with MULTOS as a test platform. A number of AE modes were implemented, however for small message sizes (up to 32 bytes), Encrypt-then-MAC (ETM) \cite{IEC200919772Encryption} was the most efficient; due to the relative speeds of the CPU and CCoP. ETM is discussed next.

\subsection{Authenticated Encryption}
\begin{figure}
\centering
\includegraphics[height=7cm, width=7cm]{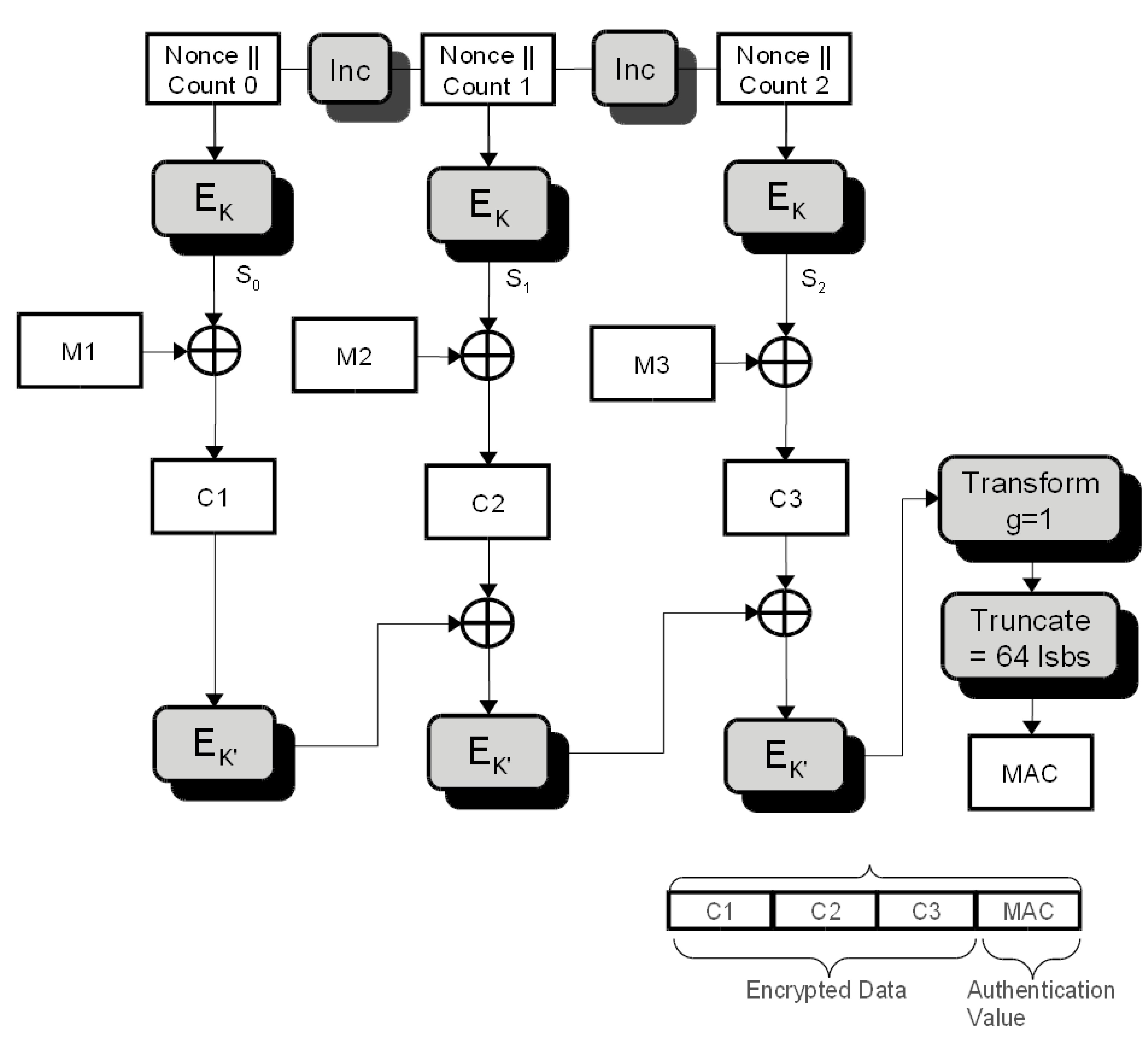}
\caption{Encrypt then MAC Authenticated Encryption}
\label{fig:etm}
\end{figure}

The ETM scheme (see Fig. \ref{fig:etm}) is \textit{mechanism 5} in ISO/IEC 19772 \cite{IEC200919772Encryption}, and is a conventional approach with separate encryption and MAC processes. The encryption stage uses block encryption in counter mode with key K, followed by a MAC (CBC\_MAC used) computation on the cipher text using a different key (K') to that used for encryption. The scheme has much to recommend it for securing grid communication. It provides authentication, confidentiality, integrity protection and has a counter for cryptographically chaining transmissions. If fixed sized data is used (e.g. the optimum 32-bytes) then the counter may be predicted in the case of lost messages, avoiding the need for re-synchronism, unless in extreme circumstances.

\subsection{Node Dynamic Initialisation}
\begin{figure}
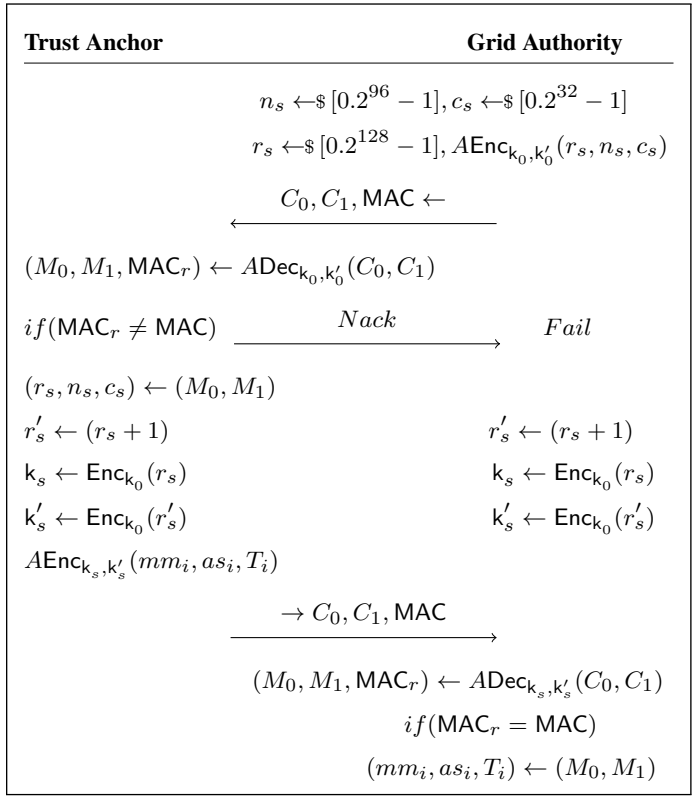

\centering
\fbox{
\pseudocode{
\textbf{Trust Anchor} \hspace{4cm} \textbf{Grid Authority} \\[0.1\baselineskip][\hline]
\<\< \\[-0.5\baselineskip]
\hspace{3.1cm}  n_s \sample [0.{2^{96}-1}], c_s \sample [0.{2^{32}-1}]\\
\hspace{3cm}  r_s \sample [0.{2^{128}-1}], A\enc_{\key_0, \key_0'}(r_s, n_s, c_s)\\
\hspace{2.5cm}  \sendmessageleft*{C_0, C_1, \mac \gets} \\
(M_0, M_1, \mac_r) \gets A\dec_{\key_0, \key_0'}(C_0, C_1) \\
if (\mac_r \neq \mac)
\sendmessageright*{Nack} Fail \\
(r_s, n_s, c_s) \gets (M_0, M_1) \\
r_s' \gets (r_{s} + 1) \hspace{4.2cm} r_s' \gets (r_{s} + 1) \\
\key_s \gets \enc_{\key_0}(r_s) \hspace{4cm} \key_s \gets \enc_{\key_0}(r_s)\\
\key_s' \gets \enc_{\key_0}(r_s') \hspace{4cm} \key_s' \gets \enc_{\key_0}(r_s')\\
A\enc_{\key_s, \key_s'}(mm_i,as_i,T_i)\\
\hspace{2.5cm} \sendmessageright* {\to C_0, C_1, \mac} \\
\hspace{3cm} (M_0, M_1, \mac_r) \gets  A\dec_{\key_s, \key_s'}(C_0, C_1)\\
\hspace{5cm}  if (\mac_r = \mac) \\
\hspace{4.5cm}  (mm_i,as_i,T_i) \gets (M_0, M_1)
}
}
\caption{Dynamic Initialisation}
\label{fig:initialisation}
\end{figure}

An SC would be initialised/personalised prior to use. This involves, for example, the setting (by a security authority) of IDs, long-term secret keys, access PINs, applications and data. Every node would be diversified, having different keys, nonces and counters, so attacking one node provides no advantage when attacking others. As the long-term secret keys are of best-practice size and do not leave the Trust-Anchor, it is acceptable to use them in normal operations, however it is even better to use them sparingly, to create session keys, which may be refreshed as necessary. The session key establishment process also provides opportunity to dynamically change nonce and counter values. The protocol steps for dynamic initialisation and session key generation are shown in Fig. \ref{fig:initialisation}, with symbols in Table \ref{table:symbols}.

The GA, begins by generating AE session values; a random number, a random nonce and a starting counter value (random or preset). These values fill two AES blocks and are AE encrypted using the long-term secret keys and the \textit{default} nonce and counter values for the particular Trust-Anchor. The resulting cipher blocks and MAC are then sent to the Trust-Anchor, which AE decrypts the message into two plain text blocks and computes its own version of the MAC. If the local and received MACs do not match, the protocol ends with a failure, otherwise the plain text is copied into the local copies of the random number, nonce and count, and the random number, and its increment are then AES encrypted under the long-term encryption key to generate the encryption and MAC session keys respectively. The Trust-Anchor then AE encrypts a dummy measurement report under session keys and sends to the GA (which also generates the keys). The GA, AE decrypts the message using session keys, and if received and locally generated MACs match, the session key establishment has succeeded. The received data is copied locally and the session keys are operational. The initialisation may be repeated, to change the AE session keys, nonce and counter values, either as security policy, or due to lost synchronism; in which case the AE count for initialisation is reset to the default for the particular Trust-Anchor.

\subsection{Data collection processing and reporting}
\begin{figure}
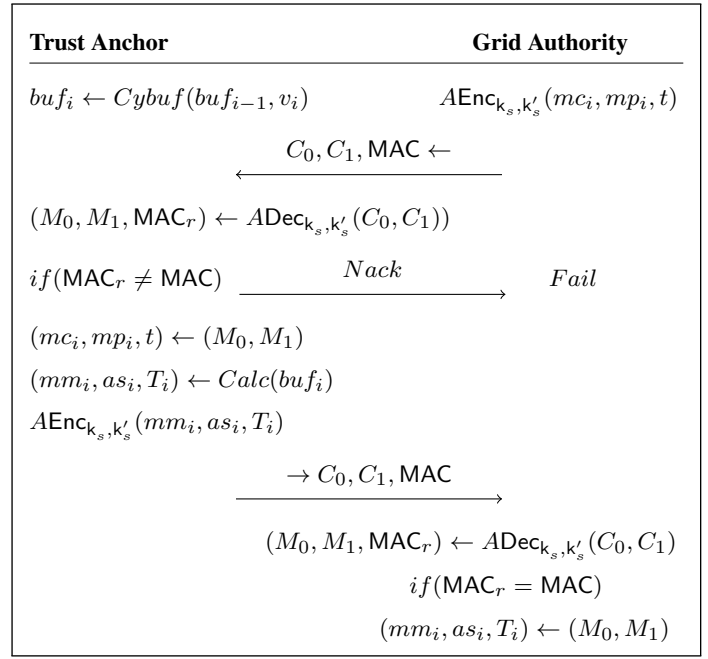

\centering
\fbox{
\pseudocode{
\textbf{Trust Anchor} \hspace{4cm} \textbf{Grid Authority} \\[0.1\baselineskip][\hline]
\<\< \\[-0.5\baselineskip]
buf_i \gets Cybuf(buf_{i-1}, v_i)  \hspace{1.7cm}  A\enc_{\key_s, \key_s'}(mc_i, mp_i, t)\\
\hspace{2.5cm}  \sendmessageleft*{C_0, C_1, \mac \gets} \\
(M_0, M_1, \mac_r) \gets A\dec_{\key_s, \key_s'}(C_0, C_1)) \\
if (\mac_r \neq \mac)
\sendmessageright*{Nack} Fail \\
(mc_i, mp_i, t) \gets (M_0, M_1)\\
(mm_i,as_i,T_i) \gets Calc(buf_i) \\
A\enc_{\key_s, \key_s'}(mm_i,as_i,T_i)\\
\hspace{2.5cm} \sendmessageright* {\to C_0, C_1, \mac} \\
\hspace{3.1cm} (M_0, M_1, \mac_r) \gets A\dec_{\key_s, \key_s'}(C_0, C_1)\\
\hspace{5cm}  if (\mac_r = \mac) \\
\hspace{4.6cm} (mm_i,as_i,T_i) \gets (M_0, M_1)
}
}
\caption{Data Collection and Reporting}
\label{fig:datacollect}
\end{figure}

Referring to Fig. \ref{fig:datacollect} and Table \ref{table:symbols}, the Trust-Anchor is continually sampling the measurement source, storing results in a cyclic buffer. The GA, begins a dialog by AE encrypting under session keys, the model control fields, the model parameters and the current time. The resulting cipher blocks and MAC are sent to the Trust-Anchor, which AE decrypts the message into two plaintext blocks and computes its own version of the MAC. If the local and received MACs do not match, the protocol ends with a failure, otherwise the plain text updates local copies of model controls, parameters and real-time. The Trust-Anchor then runs the detection algorithm on the buffered samples and calculates the model measurements, updates the alarm status, creates a Timestamp, generates a local alarm (if merited) and subsequently AE encrypts the results under the session keys and sends to the GA. If the MAC verification succeeds, the report is accepted into the modelling and state estimation application. If the Trust-Anchor does not respond, or verification fails, the GA may attempt lost message recovery.

\subsection{Lost message recovery}
A goal of the protocol is to tolerate the loss of at least two consecutive measurement reports, without loss of sample data. The reported model measurement field includes the latest sample plus up to 7 previous samples, allowing data recovery even if blocks are lost. However, the AE counter advances during encryptions/decryptions and using the wrong value will prevent correct decryption and MAC verification. As the message sizes are always the same, the counter value after a lost message can be estimated (within a small window), and so the GA can attempt AE decryption with several alternative counter values, and if successful will not lose any reported data. Similarly, at the Trust-Anchor, a MAC failure could indicate a missed or corrupted report request, and so the next count values could be tried for the AE decryption. In the case of irretrievable loss of synchronism, the GA will recover by using the dynamic initialisation process.

\begin{table}
\centering
\center
\caption{Symbol Definitions}
\label{table:symbols}
\begin{tabular}{|p{1.7cm}||p{5.5cm}|} \hline
Symbol & Description\\ \hline
${ID}$ &  Security Chip ID (128-bit, or 80-bit if ICCID))\\ \hline
${\key_0, \key_0'}$ &  Personalised long-term encryption and \mac  keys (128-bit)\\ \hline 
${\key_s, \key_s'}$ &  The current session encryption and \mac  keys (128-bit)\\ \hline
${r_s, r_s'}$ &  Random values for session key generation (128-bit)\\ \hline
${n_s}$ &  The current AE session nonce (96-bit)\\ \hline
${c_s, c_i}$ &  Starting and ith transmission, AE session counts (32-bit)\\ \hline
${M_j, C_j}$ &  The jth message and cipher blocks (128-bit)\\ \hline
${\mac_i, \mac_r}$ &  The ith AE token, and recomputed token (MAC: 64-bit)\\ \hline
${t, T_i}$ &  The time  and ith Timestamp (32-bit)\\ \hline
${as_i}$ &  The alarm status information 24-bit)\\ \hline
${dv_i, mm_i}$ &  The ith sample (15-bit) and model metrics (32-bit) \\ \hline
${mp_i, mc_i}$ &  The model parameters (208-bit) and controls (16-bit)\\ \hline
${A\enc_{\key_y, \key_y'}}$ &  Authenticated Encryption under key set y\\ \hline
${A\dec_{\key_y, \key_y'}}$ &  Authenticated Decryption under key set y\\ \hline
${\enc_{\key_y}}$ &  AES Encryption under key y\\ \hline

\hline\end{tabular}
\end{table}

\section{Event-triggered Moving Target Defence}
\label{sec:event}

\subsection{Moving Target Defence}
MTD involves using the system assets to change the underlying topology and expose FDI attacks. FDI attacks requires the attack vector be structured based on the network topology.
For real power residual, error at the individual measurement level will be the difference between the measured flows and estimated value from the system model such that real power residual can be expressed as a function of the individual meter measurements and system states by

\begin{dmath}
     r_{ij}^P = -P_{ij}^m +  V_{i}^2 g_{ij} - V_{i}V_{j} g_{ij}\cos{\Delta \theta_{ij}} \\ -V_{i}V_{j} b_{ij}\sin{\Delta \theta_{ij}}.
    \label{Pfac}
\end{dmath}

With a similar equation for resultant changes in the reactive power measured residual. If the attacker is aware of the system topology, he/she can structure the attack such that the residual is small. However, if changes to the physical system are introduced (via inductance modification in this case) and the attacker does not consider this in the updated attack vector $\textbf{z}_a$ the residual will increase as differences between the expected and injected value emerge. Where ignoring the change in voltage angles and magnitudes for a simplified view, the impact to residual will be approximated by

\begin{dmath}
     \Delta r_{ij}^P \approx V_{i}V_{j} \Delta b_{ij}\sin{\Delta \theta_{ij}} 
    \label{Pfac}
\end{dmath}

 MTD can also be enhanced to camouflage it's existence to minimise the potential for attackers circumventing MTD \cite{Tian2019EnhancedGrids} \cite{Higgins2020StealthySystems}. However, as shown in \cite{Lakshminarayana2018Cost-BenefitGrids} the cost of application will mean the system operator will want to minimise the overall use of MTD to only those times when the system is potentially under attack. As a result of this, we propose an MTD triggering scheme based on individual meter measurement deviations.  

The proposed defensive process has three main components; the anomaly detector, the active defence protocol and the final attack verification exploiting the secured measurement and communications authentication provided by the low cost MULTOS devices.
The anomaly detection can be done either locally at the distributed level or via post processing at the same place as the CSE.  
For deciding the error bounds to implement the trigger we use Holt's exponential smoothing and forecasting and have outlined this process below.

\subsection{Anomaly-detection based Triggering Strategy for MTD} 
We propose a distributed trigger (located at the individual measurement level) based on Holt-Winter's Exponentially Forecasting for anomaly detection. This distributed layer of detection confers several benefits to the grid. For one, distributed error checks can be preformed much more quickly than the CSE which can take upto a few minutes and may not even converge. Distributed measurements devices can also offer some localised control options in the case of a wider network failure of the SCADA or communications network. Crucially with respect to FDI attacks, distributed measurement based anomaly detection will not be susceptible to the same model based FDI attacks outlined in \cite{Liu2011FalseGrids}. Recent papers have shown that CSE can be attacked by structuring the attack vector in terms of the system topology i.e. $\textbf{z}_a = \textbf{h}(\textbf{x}+\textbf{c})$ where $\textbf{c}$ is a desired attack bias of length $n$. Therefore, to circumvent this type of stealthy bias injection, we consider a distributed error detection based on forecasted measurements to act as trigger for MTD. We consider the measured vector at a given bus system as discrete, linear model and follows the form:

\begin{equation} \label{residualhat}
     \hat{z}_{t+1} = F_{t}z_{t}+b_{t}+ \nu
\end{equation}

Where $\hat{z}_{t+1}$ is the estimated metered power flow measurements at a given time $t$,  $\nu$ represents the system noise which will follow a Gaussian distribution such that $\nu \hookrightarrow \mathcal{N}(0,\,Q)$. $F_t$ represents the state transition factor i.e. the expected change in the system state for a given time period and $b_t$ the state trajectory vector used to capture long run trends and will incorporate factors such as seasonality. $F_t$, $b_t$ and $\nu$ are calculated by analyzing previous data trends using Holt-Winters's exponential forecasting \cite{Taylor2007Short-termData} and contain seasonal elements so as to capture differences in intraday measurement resulting from load peaks. Anomaly detection in dynamic estimation will be the difference between the forecasted state value and the measured value for a given time and is given by

\begin{equation} \label{residual}
    e_t =  z_{t+1} -\hat{z}_{t+1}.
\end{equation}

With the assumption that residuals are independent and follow a zero mean Gaussian process. They can either be defined using static pre-set limits or more commonly using an updated variance value $\sigma^2$ from observed residuals. The introduction of this additional criteria means in order to attack stealthily (or at least not trigger MTD) the attacker would have to now satisfy both the centralised and distributed criteria such that

\begin{equation} \label{residualhat}
  \{ \textbf{z} \in \mathbb{R}^{m\times1} \wedge ||\textbf{z} - \textbf{h}(\textbf{x})||_{2} < r_c \wedge z-\hat{z}<r_d \}
\end{equation}
where $r_d$ and $r_c$ are the alarm limits for distributed and centralized triggering/detection.  
It is possible that an attacker maybe able to structure an attack in this manager but their flexibility in which loads can be overloaded will be greatly reduced. 

\subsection{Selection of Alarm Limits}
Alarm limits can be imposed considering a number of factors

\subsubsection{Type-II error}
Type-II error is an important consideration for alarm limits. It might be tempting to set limits artificially low to capture more potential attacks. However, this could result in higher unnecessary costs to the system operator as MTD will be triggered frequently when no attack persists. Frequent false errors can also reduce reasonable responses as system operator grow used to seeing a high number of alarms and lack the bandwidth to verify them efficiently.  
\subsubsection{Criticality and Capacity}
Individual measurement critically can also weigh into limit setting. A system operator may wish to lower limits for critical measurements. Conversely, areas with high additional transmission capacity, operating well below their thermal limits will suffer less consequences from overloading FDI attacks. As a result, alarm limits could be relaxed to reflect their respective innate resilience of regions and focus on other weaker areas.      

We prioritized type-II error considerations in our model and opted for limits based on the probability distribution function of a Gaussian distribution with 3 standard deviations  (SD) corresponding to a 99.7\% confidence interval (or type-II error of around 0.3\%). This is in contrast to CSE which usually operators with a 2 SD limit \cite{Liu2011FalseGrids}. We believe it appropriate to have different alarm limits for these two approaches. For one, the frequency and number of distributed checks can be much higher for distributed decision making. Where unlike CSE, there are multiple potential alarm measurements i.e. 34 alone in a 14-bus system, as opposed to CSE, which is only one centralized measurement. Also, as discussed, there is a cost in applying the MTD from an operational power flow perspective and it may not be desirable to use MTD triggering constantly when the branch power flow level offers no danger to the system. Ultimately, the selection of alarm limits comes down to engineering judgement. For highly critical regions, a system operator may tolerate increased false positives from a lower limit, in order to increase sensitivity.

In our model then the upper limit  is defined by  $ \lambda_{upper} = z_t+3\sigma$ and lower limit $\lambda_{lower} = z_t-3\sigma$ and are bound such that

\begin{equation} \label{residual}
    \lambda_{lower} < e_t < \lambda_{upper}.
\end{equation}
 For the power system models, we assume an intra-day profile similar to that of a consumer load flow. This means our forecasting model needs to approximate the underlying seasonality in creating the prediction model. The Holt-Winters method is used to capture seasonality and trend when forecasting data sets. The method is made up of a forecasting equation and accompanied by three equations for the current level $L_t$ 

\begin{equation}
   L_t = \alpha(\frac{z_t}{S_t})+(1-\alpha)(L_{t-1} + T_{t-1}).
\end{equation}

The overall trend of the data set $T_t$ described by

\begin{equation}
   T_t = \beta(L_t - L_{t-1})+(1-\beta)T_{t-1}.
\end{equation}

And also the seasonality component of the dataset $S_t$

\begin{equation}
   S_t = \gamma(\frac{z_t}{L_t})+(1-\gamma)S_{t-p}.
\end{equation}

Where $p$ is the time period in a season. $\alpha$, $\gamma$ and $\beta$ are the weighting values for level, trend and seasonality. These are selected for a minimized RMSE error based on a observing differences between the model estimates and actual values. The final predicted value using HW will be described by 

\begin{equation}
   \hat{z_t} = (L_{t-1} + T_{t-1})S_{t-p}.
\end{equation}

\subsection{Computational Considerations}

It is assumed with respect to the CSE (which is already commonly used) that operators already have the capability to perform the state estimation for their system. With regard to the computation cost of applying the Holt-Winters distributed forecasting for the event trigger, the impact is small. As shown, the basic calculations are linear and can be performed quickly even with a large number of measurements. Also, as discussed, as the method is distributed to the individual meter level the computational cost of this approach is not dependent on the size of the system i.e. we would not expect real impact from the curse of dimensionality resultant from system size as the core calculations and regressions are still dependent on individual measurements only.

\section{Experiments and Findings}
\label{sec:exp}
\begin{figure}
 \centering
 \includegraphics[height=6cm]{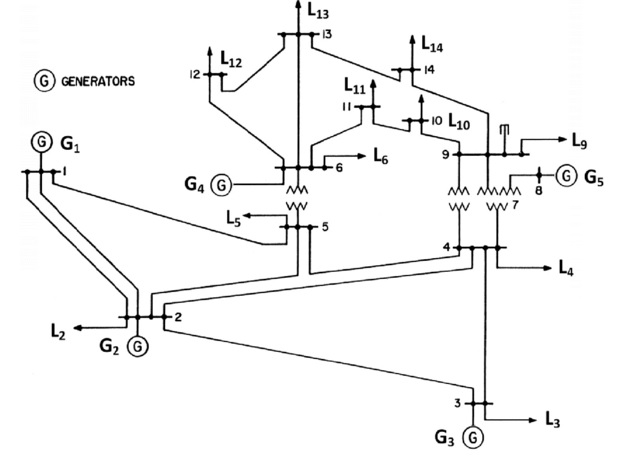}
 \caption{IEEE 14-bus system used for simulation}
 \label{fig:14bus}
\end{figure}

This section assesses the performance of the proposed detection strategy on both the standard IEEE 14-Bus test system (Fig. \ref{fig:14bus}) and the IEEE 118-Bus test system. All grid simulations were implemented using the MATPOWER toolbox in MATLAB \cite{Zimmerman2011MATPOWER:Education} and performed using Intel Core i7-7820X CPU with 64GB of ram running on a Windows 10 system.
The node performance experiments were carried out using a MULTOS Trust-Anchor development kit.

\subsection{MULTOS Trust-Anchor Performance}

The use of the MULTOS Trust-Anchor, satisfies the node attack-resistance requirements of the proposed system, however it is a CPU speed and memory restricted device, so its ability to satisfy performance requirements needs to be determined by experiment. The GA is assumed to have access to powerful server capability and so only the measurement node performance was practically investigated. The Trust-Anchor chip used for the research had digital I/O, but no Analog to Digital Converter (ADC) and so an external device was connected via its I2C bus. This did not change the attack assumptions, as it was already accepted that the analogue source value might be modified, so this just extended to the ADC chip and the I2C bus. The ADC can sample orders of magnitude faster than needed, so the effective sample rate is determined by the Trust-Anchor reading from it. Normally, the Trust-Anchor would be required to free-run, but for precise experimental measurements, it was run in command/response mode, where actions were triggered by commands from the GA.
The Trust-Anchor was used within a breakout board that allowed PC control via a USB port. GA messages were manually created and sent to the Trust-Anchor via the MULTOS MUTIL scripting utility. MUTIL logs include message timing, and a millisecond timer was also configured within the Trust-Anchor, as a calibration check.  Test software was in `C', within a single application. For message timing precision, commands were run at least 64 times before response, in order to compensate for residual measurement tolerance.

\begin{figure*}[t] 
\includegraphics[height=4.5cm, width=17.75cm]{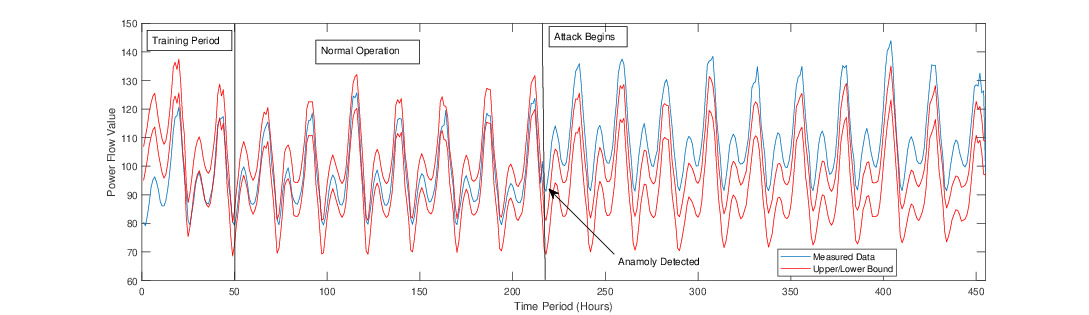}
\caption{Power flow measurement across line 1-5 in the IEEE 14-bus system over a 24 hour period. System measurements shown in blue with upper/lower bounds designated via hole-winters seasonal forecasting in red. Three time periods shown: initial training period for the forecasting, followed by under normal operation and under the FDI attack initiated. }
\label{fig:normalopdis}
\end{figure*}

The initial performance results are presented in Table \ref{table:multosbench}, entry row eight is the worst case performance test. It assumes that a GA request arrives and must first be AE decrypted. The request requires the Trust-Anchor to acquire a new sample, add it to the cyclic buffer, compute the model means, test the sample and means against the model thresholds, and finally AE encrypt the resulting report for transmission back to the GA. The total execution time is 82.83 ms, supporting a maximum repetition rate of approximately 12Hz (12/s). This suggests the Trust-Anchor could support more sophisticated, fine-grained local processing of measurement data, as part of a distributed anomaly detection system.

\begin{table}
\centering
\caption{Trust-Anchor Results}
\label{table:multosbench}
\begin{tabular}{|l|l|c|c|c|c|} \hline
& Function & Time & Blocks & Total & Max Rate\\ \hline
&  & (ms) &  & (ms) & (Hz)\\ \hline
1 & ADC Read Loop & 0.72 &  &  0.72 & 1391.30 \\ \hline
2 & As(1)+ update cyclic buf & 1.70 & &  1.70& 587.16 \\ \hline
3 & Ad(2)+ calc means & 12.88 & & 12.88 & 77.67 \\ \hline
4 & As(3)+ check thresholds & 33.52 & & 33.52 & 29.84 \\ \hline
5 & ETM decrypt & 12.20 & 2  &  24.41 &  \\ \hline
6 & ETM encrypt & 12.45 & 2 &  24.91 &  \\ \hline
7 & & & & & \\ \hline
8 & Sample Test and Report &  &  &  82.83 & 12.07 \\ \hline
\hline\end{tabular}
\end{table}

\subsection{Anomaly Detection}

As shown in \cite{NationalGrid2017ElectricityStatement}, the UK transmission network typical operates at around 80\% capacity. Given this we opted to apply a 15\% change in the system power flows at the attacker target. This would give the attacker enough flexibility to start to push these limits (at least at a regional level). In Fig. \ref{fig:normalopdis} and \ref{fig:normalopdis118} we show the anomaly detection algorithm implemented for a FDI attack line overload of 15\%. There is an initial training period for the HW seasonal forecasting for a system under normal operation. The focus is on individual line over loads and Fig. \ref{fig:normalopdis} shows line 1-5 for the 14 bus system for 400 hours of operation with a dual peak profile occurring over 24 daily measurements. In blue, we show the measured data set and red the upper and lower bounds as predicted by the Holt-winters forecast. As we can see, the first few days of implementation require a data training period have high volatility in the forecast model and high corresponding type-II error. However, this declines as the model has sufficient data to train over. Once this training period is completed, the data sets fit closely and we can see no anomalies are detected over the next week of operation past the training period. We see a similar result for the IEEE 118-bus system in Fig. \ref{fig:normalopdis118}.

In Fig. \ref{fig:normalopdis} \& Fig. \ref{fig:normalopdis118}, we initially continue basic operation for the 10 days before implementing the FDI attack. This attack is stealthy from a centralised perspective but is detected by our distributed anomaly detection. This illustrates the effectiveness of the anomaly detection which can evaluate potential FDI attacks at a distributed level (we show that these attacks would otherwise bypass CSE). It should be noted that there is potential for false positives with this kind of anomaly testing. As the forecasting is based on prior measurements if new loads are added it may take a few cycles for the forecasting to reflect the new reality. However, these type-2 error anomalies for the distributed detection should occur relatively rarely (0.3\% of measurements in a 3 SD operating window or around 5\% for a 2 SD configuration) and can be configured based on the acceptable ranges of the system operator. Furthermore, the occasional false triggering only leads to the activation of MTD to confirm the alarm, and hence the overall impact on system operation or cost is limited.         

\subsection{Event-triggered MTD}
\begin{figure*}[t] 
\includegraphics[height=4.5cm, width=17.75cm]{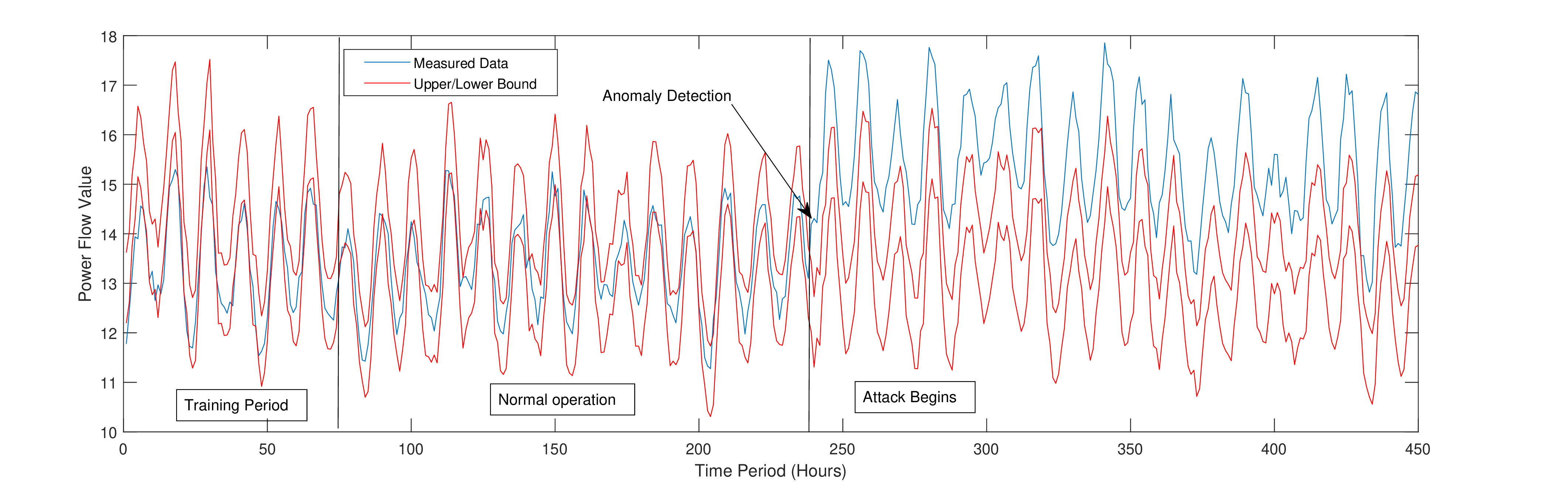}
\caption{Power flow measurement across line 1-2 in the IEEE 118-bus system over a 24 hour period. System measurements shown in blue with upper/lower bounds designated via hole-winters seasonal forecasting in red. Three time periods shown: initial training period for the forecasting, followed by under normal operation and under the FDI attack initiated. }
\label{fig:normalopdis118}
\end{figure*}

\begin{figure}
 \centering
 \includegraphics[width=8cm]{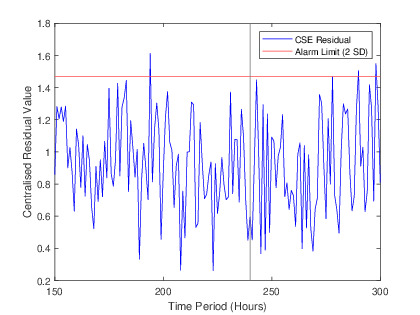}
 \caption{Residual Value for CSE under no FDI Attack for the IEEE 14-bus system. 
 }
 \label{fig:nofdiattack}
\end{figure}

\begin{figure}
\centering
\includegraphics[width=8cm]{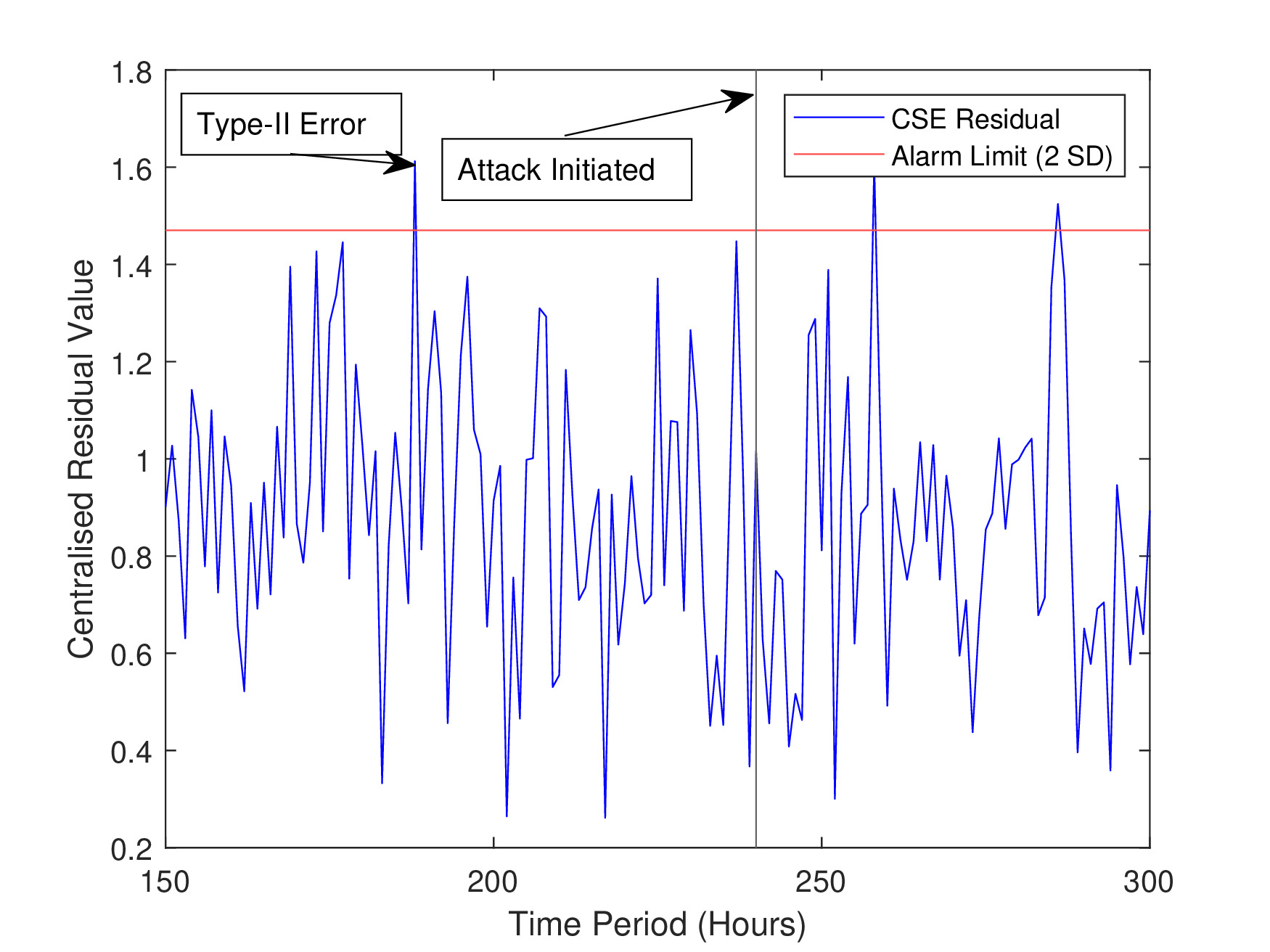}
 \caption{Residual Value for CSE under Stealthy-FDI Attack Applied from 240 hours without MTD for the IEEE 14-bus system.  
 }
 \label{fig:fdiattacknomtd}
\end{figure}

\begin{figure}
\centering
\includegraphics[width=8cm]{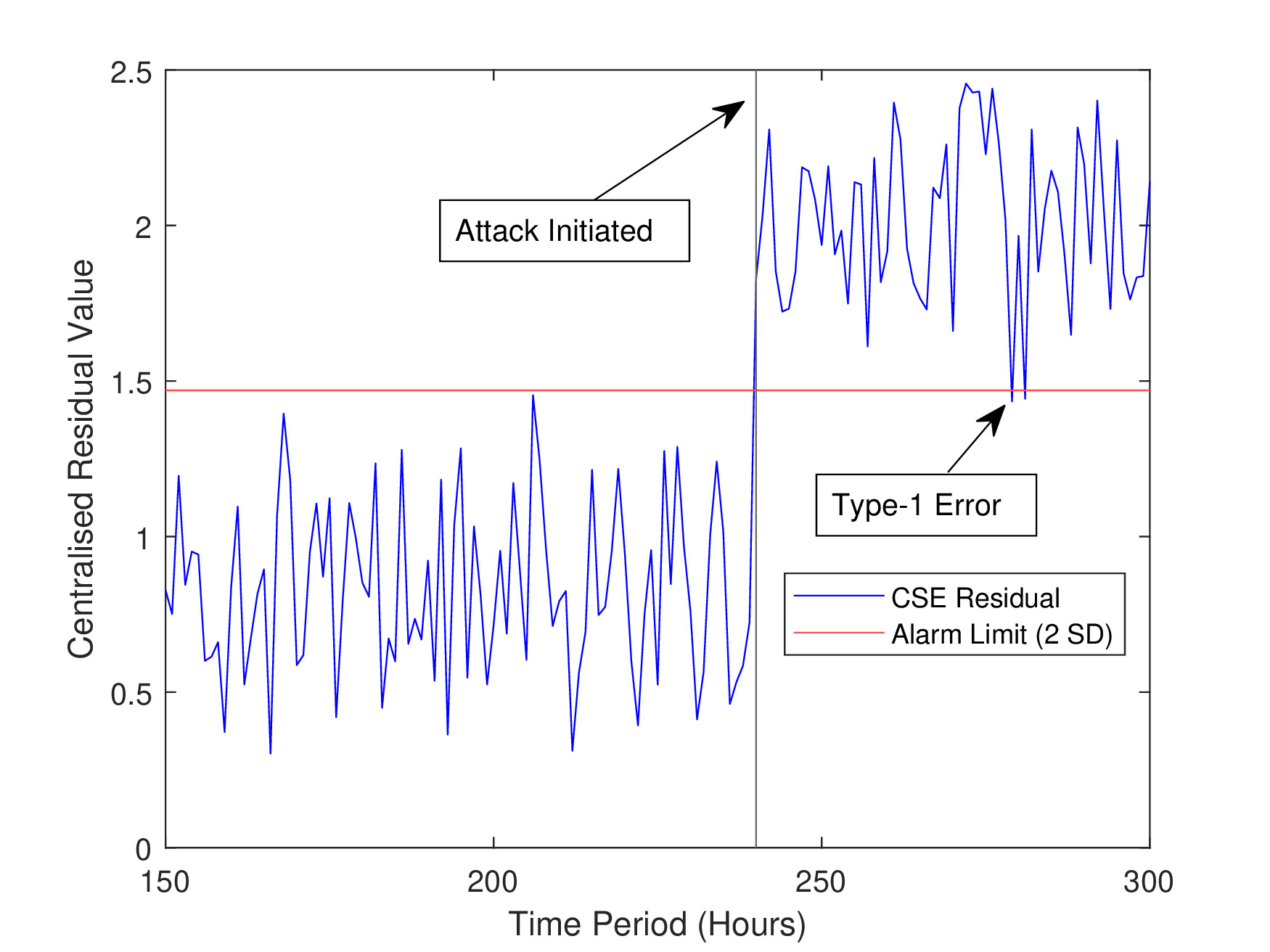}
 \caption{Residual Value for CSE under Stealthy-FDI Attack Applied from 240 hours with event triggered MTD for the IEEE 14-bus system. 
 }
 \label{fig:fdiwMTD}
\end{figure}

\begin{figure}
\centering
\includegraphics[width=8cm]{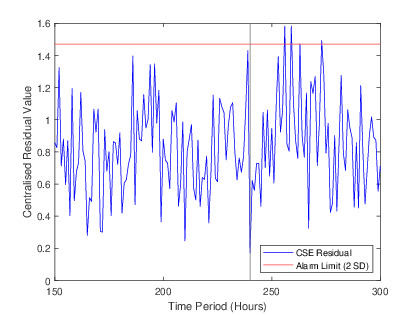}
 \caption{Residual Value for CSE under no stealthy-FDI attack with event triggered MTD triggered at 240 hours for the IEEE 14-bus system.  
 }
 \label{fig:nofdiwMTD}
\end{figure}

\begin{figure}
 \centering
 \includegraphics[width=8cm]{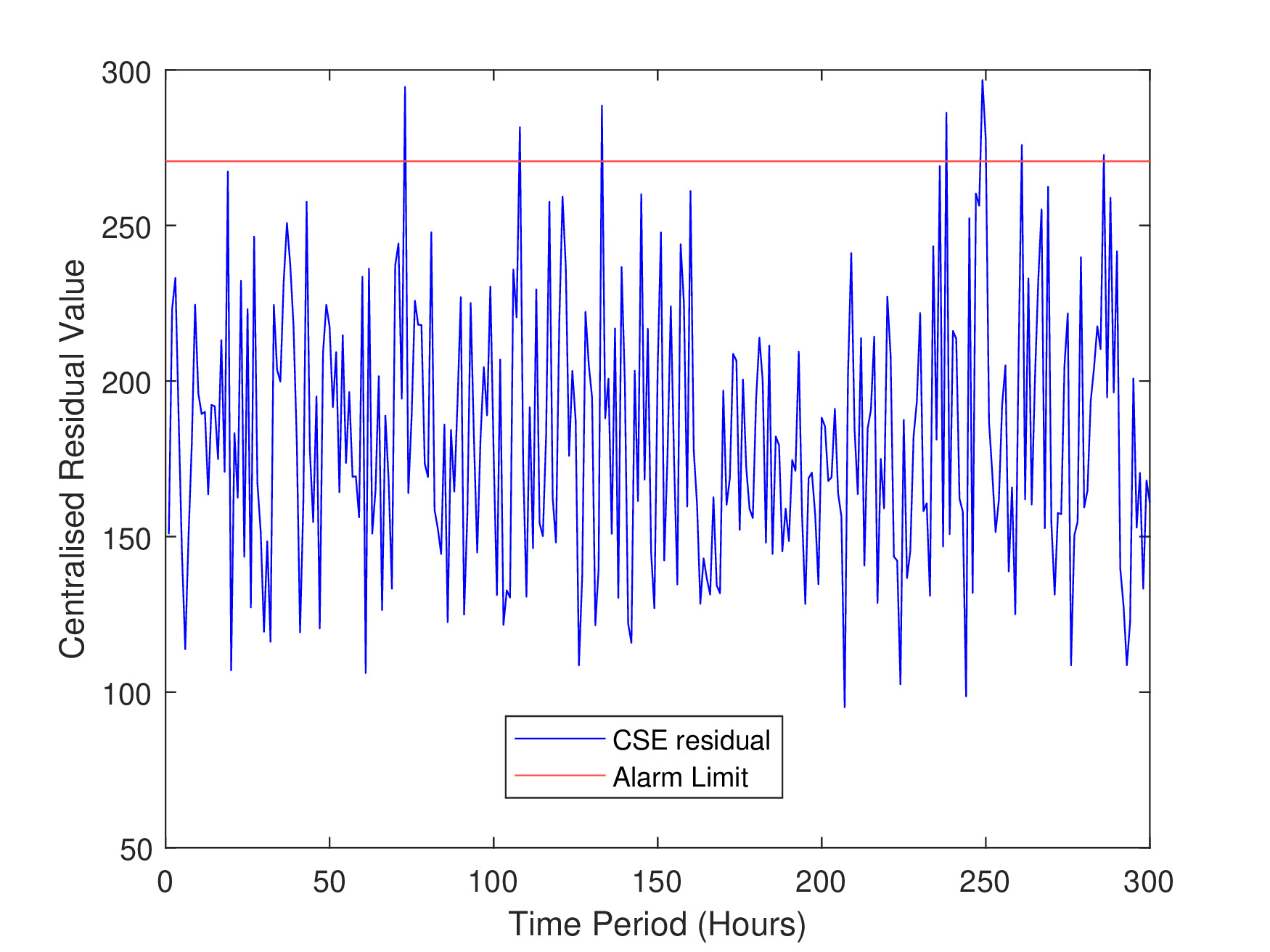}
 \caption{Residual Value for CSE under no FDI Attack for the IEEE 118-bus system. 
 }
 \label{fig:nofdiattack118}
\end{figure}

\begin{figure}
\centering
\includegraphics[width=8cm]{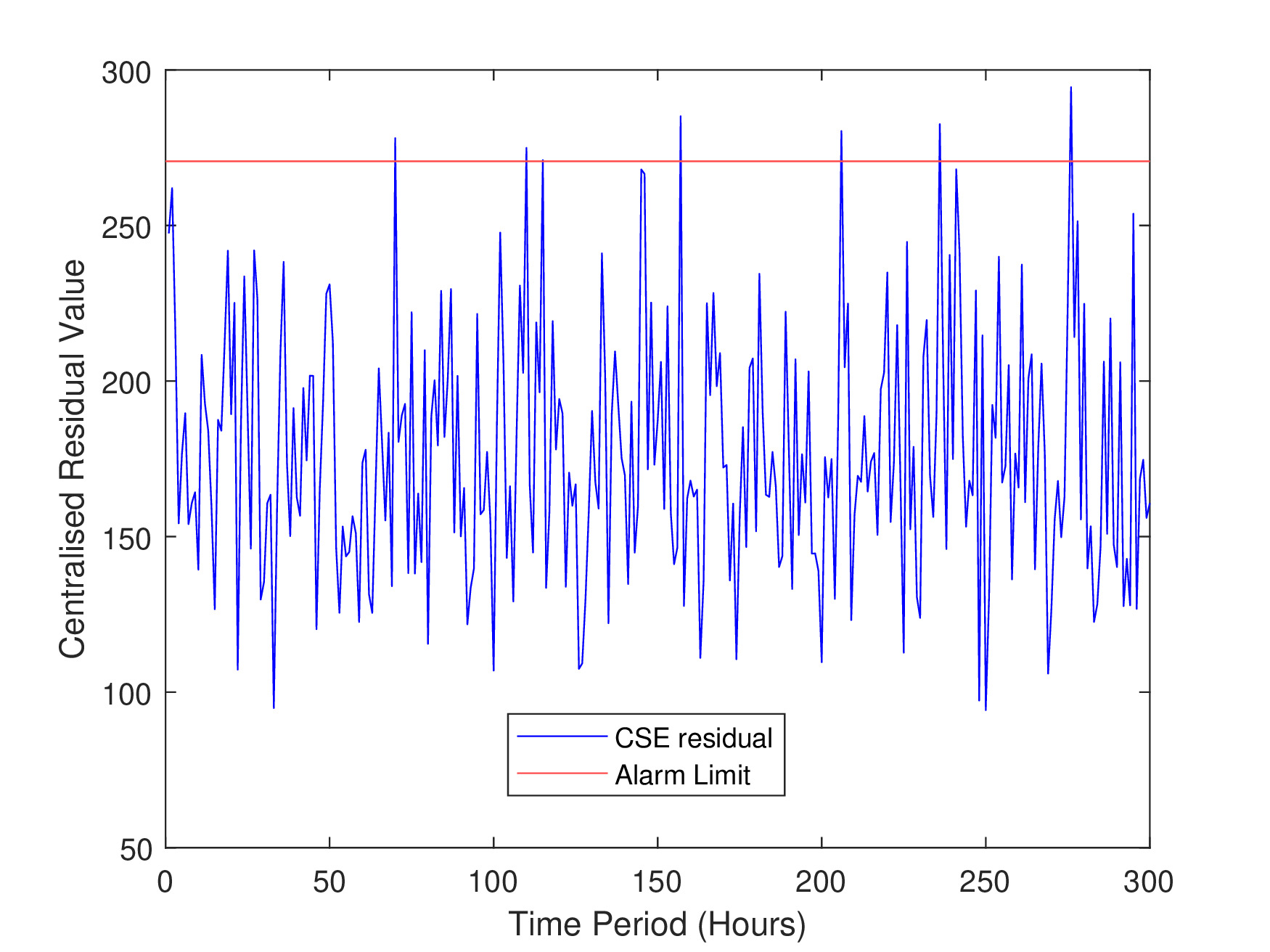}
 \caption{Residual Value for CSE under Stealthy-FDI Attack Applied from 240 hours without MTD for the IEEE 118-bus system.  
 }
 \label{fig:fdiattacknomtd118}
\end{figure}

\begin{figure}
\centering
\includegraphics[width=8cm]{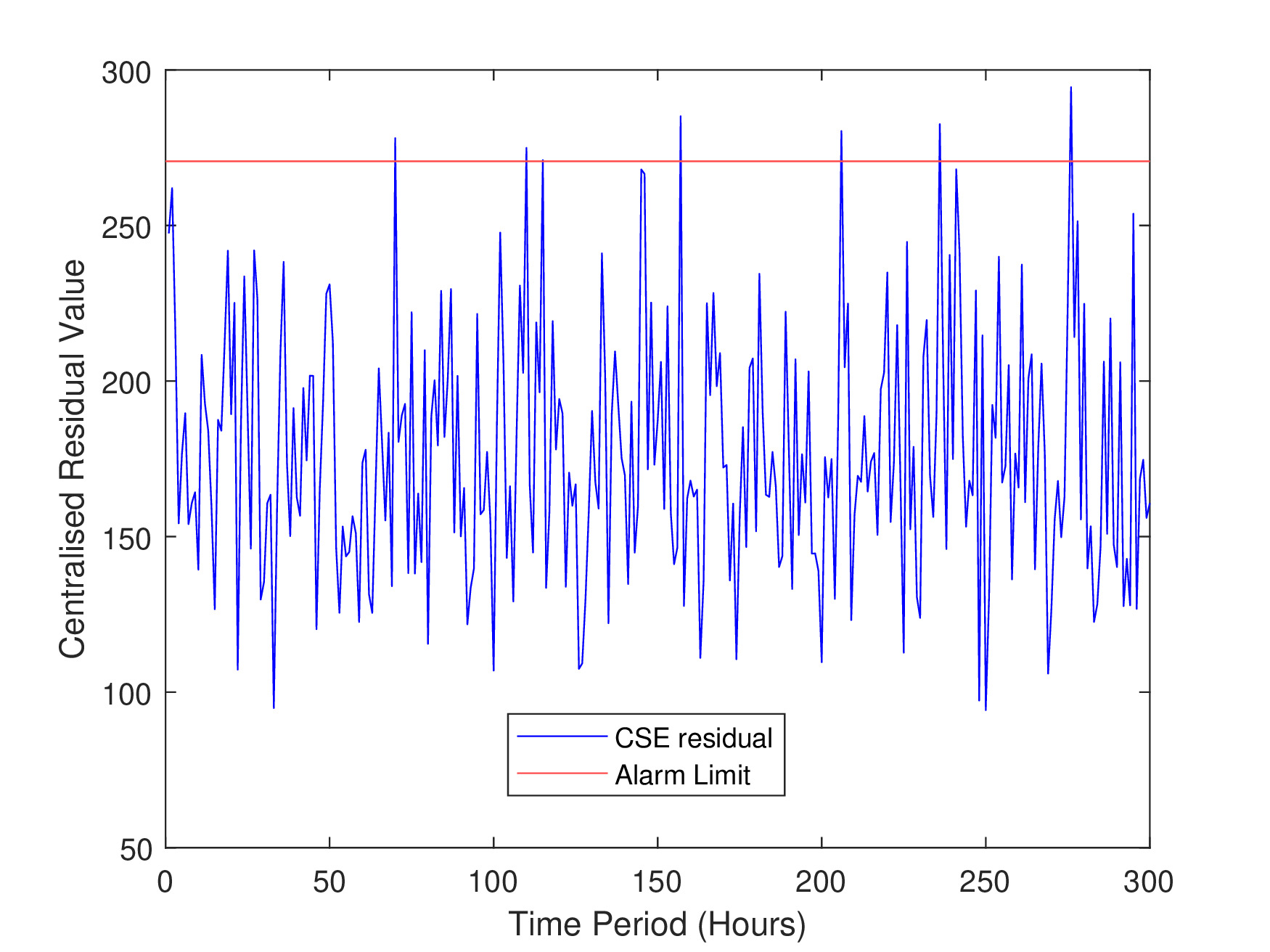}
 \caption{Residual Value for CSE under no stealthy-FDI attack with event triggered MTD implemented at 240 hours for the IEEE 118-bus system.
 }
 \label{fig:fdiwMTD118}
\end{figure}

\begin{figure}
\centering
\includegraphics[width=8cm]{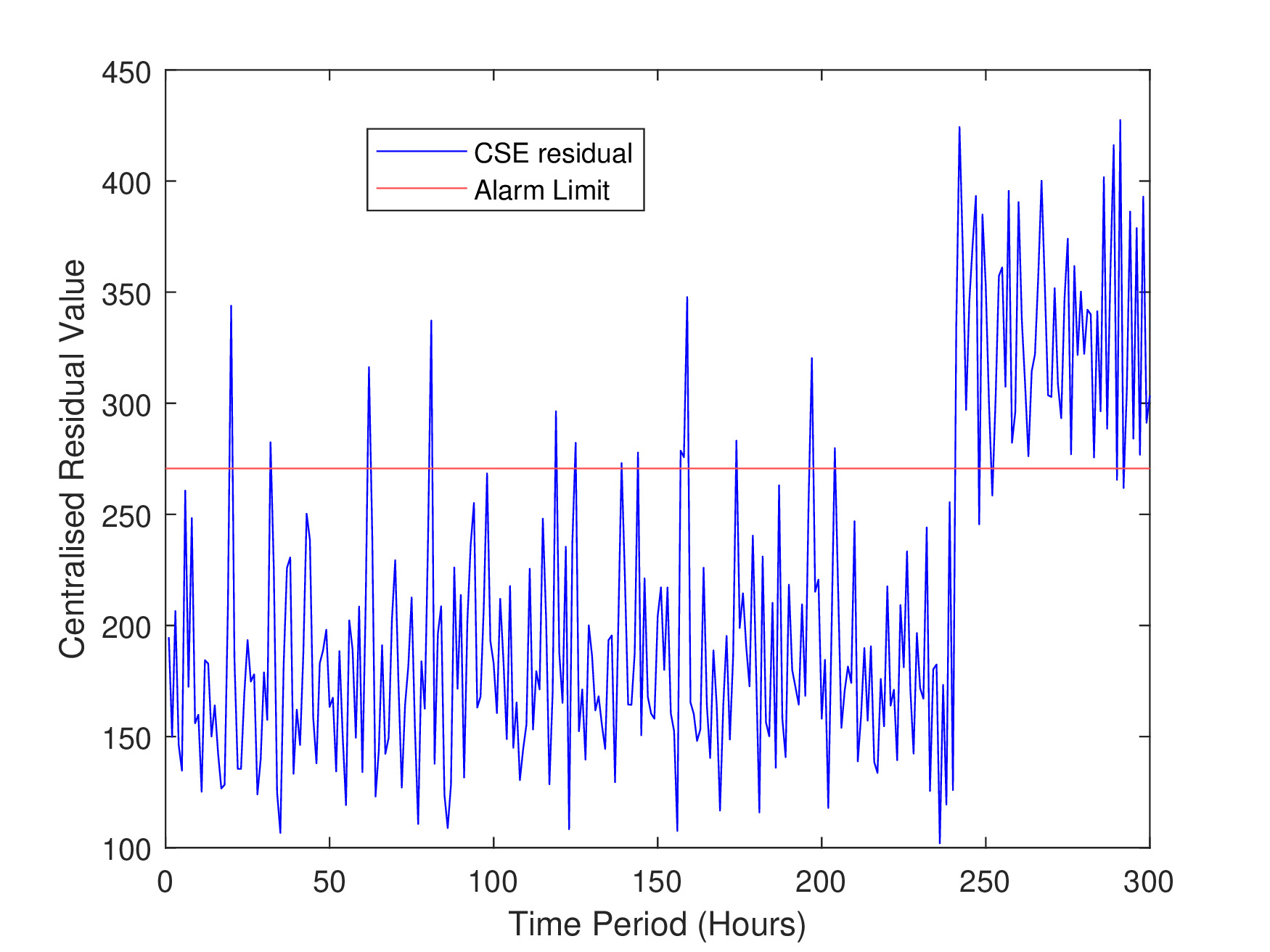}
 \caption{Residual Value for CSE under Stealthy-FDI Attack Applied from 240 hours with event triggered MTD for the IEEE 118-bus system.   
 }
 \label{fig:nofdiwMTD118}
\end{figure}

In this subsection, we demonstrate the effectiveness of the proposed event-triggered MTD. The study is carried out between the hours 150 and hour 300 as equivalent to Fig. \ref{fig:normalopdis} and Fig. \ref{fig:normalopdis118}. We first present the base case and assume no FDI attack is implemented. As shown in Fig. \ref{fig:nofdiattack} \&  \ref{fig:nofdiattack118}, the residual value of BDD in CSE under normal operation is well below the alarm limit. It should be noted  that while there is some initial type-2 error, this is not due to the FDI attack and is, in fact, consistent with 95\% chi-squared test with one anomaly point over the measurement period. In the cases where the FDI attacks are applied, we assume a stealthy FDI attack is launched from hour 240. In the case with only traditional BDD in CSE, Fig. \ref{fig:fdiattacknomtd} \& \ref{fig:fdiattacknomtd118}  shows that the residual value does not increase tangibly over the no attack case (3 type-II errors still consistent with 2 SD confidence), which means this FDI attack is not detected. However, once the proposed protocol is applied, as shown in \ref{fig:fdiwMTD} \&  \ref{fig:fdiwMTD118}, a marked increase in the overall system residual is visible, which is due to the triggered MTD, as shown in Fig. \ref{fig:normalopdis}, and the attacker's reliance on outdated model. It should also be noted that, the proposed protection protocol does not affect the centralized alarms when no attack is present as shown between hour 240 and hour 300. In the case of false triggering of MTD, \ref{fig:nofdiwMTD} \& \ref{fig:nofdiwMTD118} show that even though MTD is applied, there is no tangible increase of residual in the centralized alarms as the new set of measurement reflects the updated system physical topology. In this case, no false alarm for cyber attack will be raised.



\subsection{Security Attack Coverage}
\label{sec:plat}
Having experimentally proven that the performance of the proposed solution is well within our design requirements, we now recap on the security attack coverage, as shown in Table \ref{table:attackcoverage}. Choosing the MULTOS Trust Anchor, provides a highly attack-resistant hardware security end-point, supporting secure loading/configuration processes; making node implementation attacks \textit{practically infeasible}. The use of open/standardised algorithms and diversified credentials also removes motivation for sophisticated node implementation attacks. The use of Authenticated Encryption, prevents eavesdropping through encryption and message faking or modification via MACs. The use of a counter within the AE mode, and chaining of transmissions, prevents malicious re-ordering, and a Timestamp detects message delay. The protocol design is resilient to some lost messages (whether natural or malicious), without loss of sample data. The protocol also has resilience to loss of synchronism in counter values, and the ability to re-synch when necessary, with refresh of session keys. Normal operation cannot continue under DoS attack, however the condition is detectable by the GA and nodes; and nodes are capable of raising a local alarm when denied communication to the GA. The nodes also have the capability to continue stand-alone passive data monitoring, to detect extreme cases of source modification. Stealthy source modification is detected by the event-triggered MTD. Malicious disablement of node local alarms, can be detected by similar means.

\begin{table}

\caption{Security Attack Coverage}
\label{table:attackcoverage}

\begin{tabular}{|p{4cm}||p{0.75cm}||p{2.7cm}|} \hline
\textbf{Attack Category} & \textbf{Cover?} & \textbf{How?}\\ \hline
\hline
Node Implementation Attacks & & \\ \hline
Logical/physical/fault/side channel & Yes & Inherent in Trust-Anchor \\ \hline
Malware/personalisation/loading & Yes & Inherent in MULTOS \\ \hline
\hline
Node Communication Attacks & & \\ \hline
Eavesdropping & Yes & AE Encrypted transmissions \\ \hline
Fake message, legitimate ID & Yes & AE MAC will fail \\ \hline
Delay/replay/re-order messages & Yes & AE count chaining and Timestamp \\ \hline
Modify a legitimate message & Yes & AE MAC will fail \\ \hline
Block some legitimate messages & Yes & Protocol resilient to lost messages\\ \hline
DoS transmissions & Partly & Detect alarm at GA and nodes \\ \hline
\hline
Source Modification Attacks & & \\ \hline
Overt/extreme & Yes & Detect by passive monitoring \\ \hline
Stealthy/subtle & Yes & Detect by MTD active tests\\ \hline
\hline\end{tabular}
\end{table}

\section{Conclusions}
\label{sec:con}

This paper proposes a combined cyber-physical authentication protocol for secure and reliable, state estimation of power grids, in the presence of malicious actors. The solution combines distributed measurement nodes based on small low-power, low-cost, security chips (SC) via the MULTOS Trust-Anchor with a physical system event-triggered MTD protocol featuring distributed and centralised anomaly detection. Simulations were preformed on both the IEEE 14-bus and IEEE 118-bus systems to show the effectiveness of the proposed MTD and anomaly detection protocol in deriving FDI attacks. Practical experiments demonstrated that the Trust-Anchor could satisfy the most demanding GA request at a rate of 12/s; more than fast enough for envisaged operational scenarios. The communication protocol is resilient to missing measurement reports, without loss of data, and without the need for re-tries. Analysis of potential attacks, showed that most were countered by the choice of Trust-Anchor and the protocol design. Passive monitoring by the node (using the delegated model) detects extreme measurement source modification. DoS will disrupt operation, but is a detectable alarm event at the nodes and the GA; and a node's passive measurement monitoring will continue under DoS attack. On top of these communications protections, the physical protection protocol, looks for distributed anomalies and probes for stealthy FDI attacks using grid system assets.

\subsection{Future Work}
The vast majority of works in the field of FDI attacks have been based purely in simulation. Clearly, there is room for test-bed experiments for FDI attacks to observe practical examples with a view to implementation of protections into the grid. Also, on the physical side, expansion into different type of MTD protocols i.e. watermarking might also provide an interesting avenue for research. From a cyber perspective, areas such as the effect of high-rate sampling of measurement sources to detect identifying characteristics, for early warning of potential problems, or subtle attacks; invisible to current systems would also be of interest.

\section*{Acknowledgment}

This work was partly supported by ESRC grant (ES/T000112/1) and EPSRC Centre for Doctoral Training in Future Power Networks and Smart Grids (EP/L015471/1). The authors would like to thank Chris Torr at MULTOS for his expert support and guidance, and Crisp Telecom Ltd. for access to test and development facilities.

\bibliographystyle{iet.bst}

\end{document}